\documentclass[12pt,a4paper]{article}

\usepackage[utf8]{inputenc}
\usepackage[T1]{fontenc}
\usepackage[english]{babel}
\usepackage{graphicx}
\usepackage{amsmath}
\usepackage{hyperref}
\usepackage{authblk}
\usepackage[backend=bibtex,style=numeric]{biblatex} % Usa biblatex per la gestione della bibliografia
\usepackage{geometry}
\usepackage{array}

\usepackage{algorithm}
\usepackage{algpseudocode}

\usepackage{subcaption}

\title{Convolutional Neural Network Design and Evaluation for Real-Time Multivariate Time Series Fault Detection in Spacecraft Attitude Sensors}

\newcommand{\authorEmail}{riccardo.gallon@airbus.com}

\author[1,2]{Riccardo Gallon\thanks{\authorEmail}}
\author[1]{Fabian Schiemenz}
\author[2]{Alessandra Menicucci}
\author[2]{Eberhard Gill}

\affil[1]{\footnotesize\textit{Airbus Defence and Space GmbH, Claude-Dornier Stra\ss e, Immenstaad am Bodensee,
88090, Germany.}}
\affil[2]{\footnotesize\textit{Department of Space Systems Engineering, Faculty of Aerospace Engineering, Delft University of Technology, Kluyverweg 1, HS Delft 2629, Netherlands.}}

\date{}

\begin{document}

\maketitle

% EDIT HERE
% Please add your abstract here, i.e., \begin{abstract}<Your abstract text>\end{abstract}
\begin{abstract}
%%%
Traditional anomaly detection techniques onboard satellites are based on reliable, yet limited, thresholding mechanisms which are designed to monitor univariate signals and trigger recovery actions according to specific European Cooperation for Space Standardization (ECSS) standards. However, Artificial Intelligence-based Fault Detection, Isolation and Recovery (FDIR) solutions have recently raised with the prospect to overcome the limitations of these standard methods, expanding the range of detectable failures and improving response times. \\
This paper presents a novel approach to detecting stuck values within the Accelerometer and Inertial Measurement Unit of a drone-like spacecraft for the exploration of Small Solar System Bodies (SSSB), leveraging a multi-channel Convolutional Neural Network (CNN) to perform multi-target classification and independently detect faults in the sensors.\\
Significant attention has been dedicated to ensuring the compatibility of the algorithm within the onboard FDIR system, representing a step forward to the in-orbit validation of a technology that remains experimental until its robustness is thoroughly proven. An integration methodology is proposed to enable the network to effectively detect anomalies and trigger recovery actions at the system level. The detection performances and the capability of the algorithm in reaction triggering are evaluated employing a set of custom-defined detection and system metrics, showing the outstanding performances of the algorithm in performing its FDIR task.
%%%%
\end{abstract}

% EDIT HERE
% The main document. Please add your content as desired.
% We provide examples for adding figures, equations, and tables. Please stick to the style used in this template.
\section{Introduction} \label{sec:intro}
Failure Detection, Isolation, and Recovery (FDIR) is critical in every operational spacecraft, as undetected failures can have severe consequences, potentially jeopardizing mission uptime. Traditional FDIR approaches typically employ straightforward thresholding techniques on onboard parameters. While these methods offer robustness, they may lack efficiency in detecting certain types of failures. Operational FDIR systems in European satellites are based on the so-called Packet Utilization Standard (PUS, \cite{ECSS-E-ST-70-41C}), which defines a number of different Services regulating space-ground communications. Among these, Service 12 (Onboard Monitoring), Service 5 (Event Reporting), Service 19 (Event-Action), and action-type services like Service 18 (Onboard Control Procedure) or 21 (Request Sequencing) are mostly used in FDIR. \\
Service 12 (S12) is responsible for onboard parameter and functional monitoring (pMon and fMon) for failure detection purposes. The monitoring can apply to the value of a signal, its expected value or its variation, checking instances where they fall Out-of-Limit (OOL) beyond intervals defined during mission design. Typically, a logical combination (AND/OR) of pMons is associated to an fMon, which triggers upon the associated pMons registering OOL for a specified number of consecutive samples over a defined time interval (\textit{persistency}). Service 5 is employed to link the triggered fMon to an Event ID, which usually represents the cause of the detected failure. Service 19 links the Event ID to a recovery action. Finally, action-type services (e.g. S18 or S21) are used to execute the recovery actions.\\
The FDIR subsystem typically includes multiple chains of the above-mentioned PUS Services, structured across five hierarchical levels defining how to address failures. The levels go from unit to subsystem to system level and can be activated by the adjacent lower level only when the latter cannot resolve or isolate a fault.\\
Although representing the operational state of the art, PUS-based FDIR presents several well-known limitations, which mainly relate to its specific design for univariate time series and to the parameter monitoring mechanism. On the one hand, failures impacting several telemetry signals coming from different components and/or subsystems, as well as inherently multivariate data (e.g images) cannot efficiently be analyzed through PUS-based FDIR because they do not fit into the univariate time series definition. On the other hand, failures affecting univariate time series where the signal evolves anomalously within the boundaries set by the thresholds exist, and are traditionally undetectable unless their consequences become visible to the PUS Services on higher levels. Usually, this means the faults propagated and increased in severity.\\

Stuck values in multivariate time series data are a typical example of faults where the PUS-based detection shows inefficiency. These faults cause the signal to remain stuck on the same value over time, either involving one or all the components, eventually not exceeding the threshold boundaries. The most common detection methodology for stuck values is based on the consequences that they cause in other parameters of the system, eventually falling OOL, and are detected through higher-level FDIR. This approach, while representing the operational state of the art, lets the fault evolve uncontrollably for a certain time, thus constitutes a limitation which is worth trying to mitigate or solve. The application of Artificial Intelligence (AI) algorithms to the problem of stuck values arises from the need to enhance the detection of these faults, especially in terms of early detection and accuracy. To this purpose, this work employs the use case of the \textit{Astrone KI} spacecraft to provide data and demonstrate the validity of the presented approach.\\ \textit{Astrone KI} is a drone-like spacecraft designed for autonomous flight within the challenging environment of an SSSB, including take-off and landing. The distinctive mission profile motivates the selection of an accelerometer and an Inertial Measurement Unit (IMU), deliberately kept separate to measure both accelerations and angular rates. This setup aims to achieve precise attitude determination, particularly when coupled with cameras and LiDARs, that also serve the broader purpose of enabling vision-based navigation and guiding the drone's relocation maneuvers on the SSSB's surface. The \textit{Astrone KI} system is meant to provide the multivariate time series for the detection of stuck values by means of the mentioned accelerometer and IMU. Most importantly, the infrastructure given by the spacecraft allows to investigate the AI-based FDIR functionalities on a realistic dataset, while enabling the AI integration within the PUS-based FDIR.\\ 

The contribution of this work to the state of the art is three-fold. First, a multi-target, multi-channel, CNN-based approach to fault detection in time series coming from the accelerometer and IMU mounted onboard the vehicle is proposed. Second, a procedure to integrate the algorithm in the framework of a functional simulator of the spacecraft is described, facing the issues connected to online fault detection, defining the pipeline from input pre-processing to the classification of anomalous signal samples, concluding with the recovery action triggering by the PUS-based FDIR chain implemented onboard. Lastly, a set of case-specific metrics is proposed, in order to evaluate the AI algorithm as integrated within the onboard FDIR subsystem, capable of expressing on one side the capability of the algorithm to detect faults, on the other side the performances of the whole AI-based FDIR.\\

This paper is organized as follows: Section \ref{sec:background} presents a review of the related works on AI-based FDIR for time series. Section \ref{subs:datagen} details the process of dataset generation and anomaly injection for the purpose of training and testing the neural network. Section \ref{sec:FD} presents the FDIR methodology, including data manipulation, CNN algorithm utilization, and a specially designed integration strategy with the onboard system. Section \ref{sec:perf_ev} delineates the performance metrics tailored for the solution presented in Section \ref{sec:FD}, along with the optimization process carried out to refine these metrics and the results obtained. Section \ref{sec:results} presents the conclusions of the paper and an outlook on future work.

\section{Background} \label{sec:background}

AI-based onboard FDIR, especially about time series, is a recent trend, but it has been already identified as one of the most promising fields of research due to the foreseen benefits that AI may bring to this specific functionality (\cite{mess2019techniques}, \cite{murphy2021machine}, \cite{russo2022using}, \cite{ciancarelli2023innovative}). More specifically, scientific literature concentrates more on the field of Failure Detection and Isolation than on the Recovery part, which typically relies on conventional methods (e.g. configuration change, power-cycle, redundancy switch). A variety of AI methodologies are employed to address the FDI task, where Deep Learning (DL) is the recently-emerged alternative to the more traditional Machine Learning (ML).\\
ML applications to anomaly detection in time series include mainly Support Vector Machines (SVM, \cite{xiong2011anomaly}, \cite{gao2012fault}), Clustering Methods (\cite{liu2017satellite}, \cite{azevedo2012applying}, \cite{gao2012unsupervised}) and Decision Trees (\cite{kuhn2013decision}). These approaches, together with their very well-established state in academic literature, are also the ones that meet the desired interpretability features in the onboard FDIR. They are indeed based on fully deterministic algorithms, in contrast to Deep Learning algorithms which are essentially black boxes.\\
An example of ML-based FDIR is provided in \cite{xiong2011anomaly} and \cite{gao2012fault}, which present the application of an SVM for anomaly detection in real telemetry data including voltages, currents and temperatures. A first pre-processing step is carried out by expressing data in terms of statistical descriptors and applying Principal Component Analysis (PCA). Following, the anomaly detection is performed by the SVM. Clearly, the classifiers SVMs in \cite{xiong2011anomaly} and \cite{gao2012fault} are guided by the choice of the data features given as input, which on the one hand allows to interpret the reasoning of the algorithm, on the other hand excludes possibly better solutions. Related to this, the main reason why DL has replaced ML during the years is that the former, being a black-box model, is able to grasp more complex features of the input data and establish relations among them. Indeed, it has proved itself very efficient in anomaly detection tasks, especially dealing with multi-dimensional data such as images (\cite{wang2021unified}) or multivariate time series (\cite{li2023deep}, \cite{lakey2024comparison}), as it will be clarified in the following examples. \\

Studies on DL-based anomaly detection for time series can be divided in two main approaches: direct classification of the input or its reconstruction.\\
Direct classification applications come mainly from outside the space domain, with this not being an obstacle to their applicability from the perspective of the algorithm only, but instead a resource to provide potentially adaptable ideas. In the case of \cite{cui2016multi}, the authors provide a multi-channel CNN capable of classifying a time series by feeding each channel with a differently manipulated version of the series itself, in order to orient the feature recognition and boost the successive classification stage. The solution has been tested on 46 datasets coming from different real-world domains and compared to 14 other state-of-the-art classifiers, proving its outstanding performances. Similarly, \cite{canizo2019multi} employ mixed Convolutional-Recurrent Neural Network architectures for time series classification, where a comparison is established between 15 different networks including the multi-channel and multi-head cases, as well as different kinds of Convolutional and Recurrent layers. The test dataset is a two-classes dataset drawn from an industrial framework and dealing, among the others, with linear accelerations, angular rates and voltages. \\
A hybrid approach between ML and DL, related to the space domain, is presented in \cite{mansell2021deep}, where a One-Class Support Vector Machine (SVM) followed by a Long Short-Term Memory (LSTM) neural network performs fault detection and isolation on raw telemetry data coming from an Attitude Determination and Control System (ADCS) simulator inherited from the mission LightSail 2. Here, 21 different kinds of faults at unit level are applied to onboard sensors and actuators, in order to test the FDI capability. The solution relies firstly on the discrimination made by the SVM and some additional rule-based checks to perform the detection task, then complemented with isolation by the LSTM network. Such architecture performs almost perfectly in accuracy on the majority of the injected faults and constitutes a pioneering work on AI-based FDIR in the space domain.

Signal reconstruction methodologies aim typically at predicting the value of time series forward in time. Specifically to FDIR, they are usually complemented by classifiers which make use of the predicted (\textit{reconstructed}) signal to perform the detection and eventually identification tasks (\cite{hundman2018detecting}, \cite{li2019stacked}, \cite{baireddy2021spacecraft}, \cite{xiang2021robust}). In \cite{hundman2018detecting}, LSTM-based predictors with dynamic thresholding mechanisms are used to perform anomaly detection tasks and are shown to obtain state-of-the-art results on datasets coming from telemetry of well-known NASA satellites (e.g. Mars Reconnaissance Orbiter). The technique is based on applying a sliding window to the dataset made of time series and feeding it to the network, whose output is a reconstructed version of the input window itself. Then, the proper anomaly detection step consists in taking the difference between the output of the network and the real telemetry (\textit{reconstruction error}) and comparing it to a threshold, chosen based on the probability distribution of the data in the specific window (i.e. \textit{dynamic thresholding}).\\
\cite{li2019stacked} took over the work of \cite{hundman2018detecting} introducing a stacked predictor including LSTM layers and an SVM, which was able to perform on the same dataset.\\
A more recent approach is presented by \cite{baireddy2021spacecraft}, which employ transfer learning techniques to improve the work of \cite{li2019stacked} and \cite{hundman2018detecting}, using again the same dataset but varying the predictor shape. In this case, the predictor is indeed based on LSTM and built first as a general predictor for time series, then fine-tuned employing weights transfer to better fit the training data. Finally, a dynamical thresholding mechanism inherited from \cite{hundman2018detecting} is applied for the classification stage. The solution presented in \cite{baireddy2021spacecraft} proves to obtain better performances than both \cite{li2019stacked} and \cite{hundman2018detecting}, as well as other state-of-the-art methods based on the same predictor-classifier approach. Plus, it introduces transfer learning in the AI-based FDI, marking an emerging trend in the field-specific academic literature (\cite{di2021ensemble}, \cite{baireddy2021spacecraft}, \cite{zhang2023anomaly}). This methodology is indeed capable to mitigate abrupt alterations in model structures when training for closely related tasks, and consequently to reduce the amount of training data needed, particularly in scenarios where those are limited, e.g. real telemetry.\\

Techniques of direct classification are typically less employed than signal reconstruction ones in the space domain, because they allow less flexibility in the classification phase. Indeed, while in the former case the algorithm directly outputs the label associated to the input (i.e. \textit{fault}/\textit{no fault}), the latter outputs in a wide sense the input itself, which has to be classified afterwards by complementary methodologies. In other words, the black-box part of the model goes from input to output in the case of direct classification, while it is limited to the signal reconstruction in the other case, which can instead employ robust and/or interpretable classifiers. \\
While the signal reconstruction approach may seem more suitable to FDIR applications, it needs specific attention in the integration phase of this specific study, where the match between the reconstruction mechanism and the persistency-reaction triggering mechanism of PUS mandates further investigation. This is not the case for direct classification methodologies which mainly present the challenge of where to place the algorithm in the PUS chain. For this reason, the present research concentrates on the direct classification, although not excluding the signal reconstruction for future work.\\

Concerning the evaluation of ML-based anomaly detection algorithms with case-specific metrics, the problem consists in defining new ways of evaluating the performances of an AI algorithm in those fields of application where conventional metrics (e.g. Precision, Recall, F1-Score) are not representative. Specifically to failure detection tasks for time series, where faults are more likely to occur in intervals than in single samples, a measure of the overlap of these intervals with the AI prediction can give much better insight in the detection than measuring sample-by-sample metrics (e.g. accuracy).\\
\cite{mathonsi2022multivariate} and \cite{nalepa2022evaluating} deal with this performance evaluation topic applied to predictor-classifier architectures. More specifically, both works establish a metric about the time interval between prediction and failure occurrence, named \textit{early-detection score}, which corresponds to how early is it possible to predict the fault in their predictor case. In addition, \cite{nalepa2022evaluating} introduce the \textit{Dice coefficient} to measure the overlapping between the predicted signal and the actual one, as well as a new definition of True Positives based on the predicted and actual overlapping intervals.

To the best extent of the author's knowledge, none of the techniques exposed about time series anomaly detection has been integrated to jointly operate within the onboard FDIR subsystem, especially the European PUS-based FDIR as it is exposed in Section \ref{sec:intro}, leaving a consistent gap in the state of the art. The gap mainly concerns how to interface the AI output with the rest of the PUS chain, which means assessing the capability of the AI to perform Failure Detection, Isolation or ultimately include Recovery. Interfacing the AI and the PUS-based FDIR is a crucial step of the present research, together with the more general system engineering aspects of setting requirements to the AI design and functionalities.\\
Furthermore, while the work of \cite{nalepa2022evaluating} and \cite{mathonsi2022multivariate} presents intervals-based evaluation of their predictor-classifier algorithms, another gap in the state of the art is identified in the application of this evaluation approach to direct-classification algorithms, and how it can be made specific for the case of integration alongside PUS-based FDIR.

\section{Fault Injection and Dataset Generation} \label{subs:datagen}

\begin{figure*}[h!]
    \centering
    \includegraphics[width=0.45\linewidth]{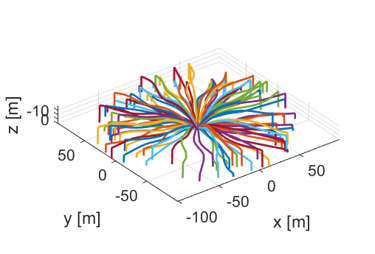}
    \caption{Randomly Generated Trajectories}
    \label{fig:rand_traj}
\end{figure*}

In the present study, the generation of the dataset to be subject to anomaly detection utilizes components of the AOCS Offline Simulation Environment (AOSE, \cite{wiedermannmodel}) employed in the Astrone KI project, developed within Airbus Defence and Space GmbH. This simulator includes proprietary models of sensors, actuators, dynamics, attitude determination and control algorithms, as well as a PUS-based FDIR module which is used to test the proposed AI-based FDIR functionalities.\\
The unit models of the accelerometer and IMU are simulated in AOSE providing as input the spacecraft position and velocity over predefined trajectories and commanding the fault injection based on specific requirements. \\
Spacecraft position and velocity are obtained simulating random monodirectional trajectories, equally spaced over the yaw angle and departing from the same starting point, assuming a hypothetical flat asteroid surface (Figure \ref{fig:rand_traj}).

\begin{figure*}[h!]
    \centering
    \includegraphics[width=\linewidth]{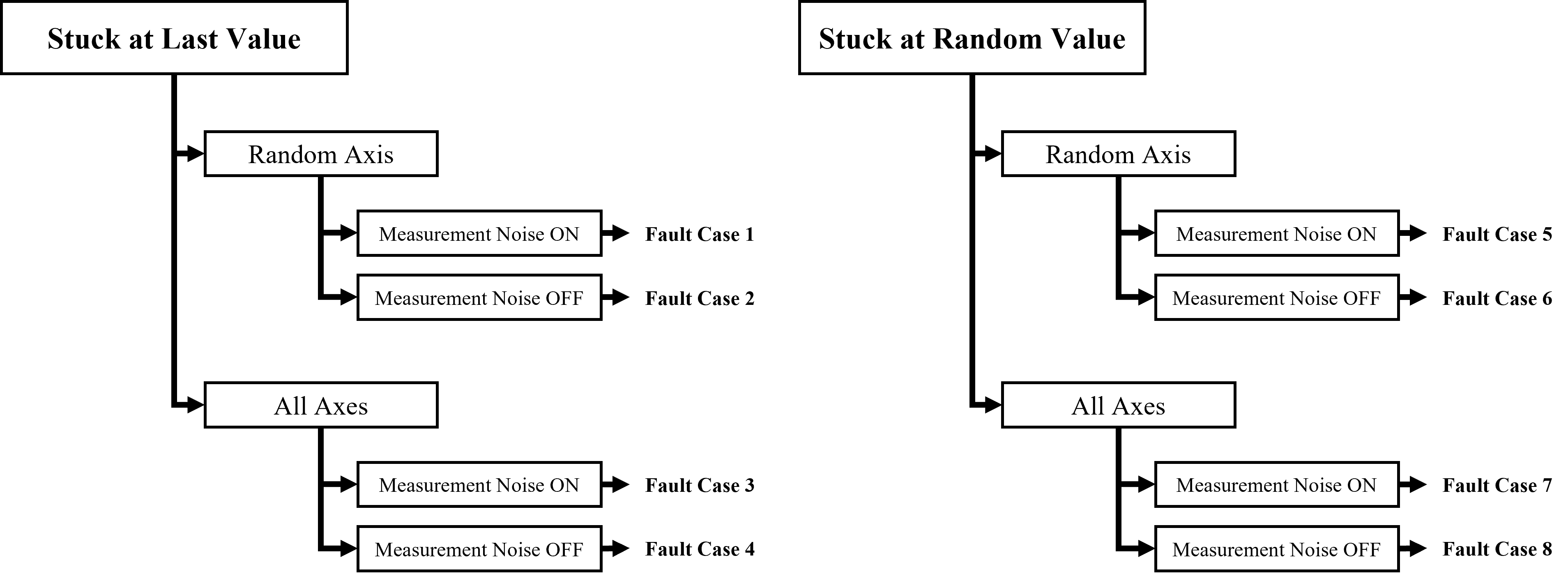}
    \caption{Injected Fault Cases}
    \label{fig:inj_fault}
\end{figure*}

The fault injection is commanded by a vector of bits establishing both the fault to inject and the injection time sample. Injected faults comprise two main cases: \textit{Stuck at last value} and \textit{Stuck at random value}, where \textit{last value} is the one assumed by the signal at the fault occurrence, while \textit{random value} is any other value but the last one, either within or beyond the nominal range. Targeted fault cases are also distinguished by axis occurrence and presence or absence of measurement noise on top of the fault (Figure \ref{fig:inj_fault}).

The process of dataset building has been oriented by a set of requirements to guide the subsequent learning of the network. The established requirements act on the injected faults and include: randomized time sample of occurrence, specific boundaries of duration of the faults and interval of time distance between subsequent faults. The randomized time sample for fault occurrence is representative of the possibility of the fault to occur at any time during operations. The lower bound to the fault duration is set based on expert considerations on realistic duration of stuck values, balancing the need to give time to the AI to recognize the fault and trigger a PUS-based recovery. The upper bound is also empirically set to be able to insert a high number of faults in the simulations, in order to focus the AI learning on early fault detection. The time interval to separate two successive occurrences of the faults is set within the same sensor and across both ones, in order to be compliant with realistic faulty situations: on the one hand, the time for a previous recovery action to occur in the single sensor shall be kept into account, on the other hand it is very unlikely that two faults occur simultaneously in both sensors. Table \ref{tab:fail_inj_par} shows the numerical values of the fault injection parameters. 
\begin{table*}[hbt!]
\centering
\caption{\label{tab:fail_inj_par} Fault Injection Parameters}
\begin{tabular}{|l|l|}
\hline
    Parameter                                 & Value (samples) \\ \hline
    Lower Bound Fault Duration                & 30              \\ \hline
    Upper Bound Fault Duration                & 110             \\ \hline
    Distance Between Subsequent Faults        & 305             \\ \hline
\end{tabular}
\end{table*}

The \textit{Upper Bound Fault Duration} and \textit{Distance Between Subsequent Faults} in Table \ref{tab:fail_inj_par} are also constrained by the \textit{Widows Length} parameter in Table \ref{tab:hp}, derived from the optimization process described in Section \ref{subs:opt}. This constraint will be further detailed after a better description of the CNN architecture is provided.

\section{Fault Detection} \label{sec:FD}
\subsection{Data Analysis} \label{subs:data_analysis}

\begin{figure*}[h!]
    \centering
    \includegraphics[width=\textwidth]{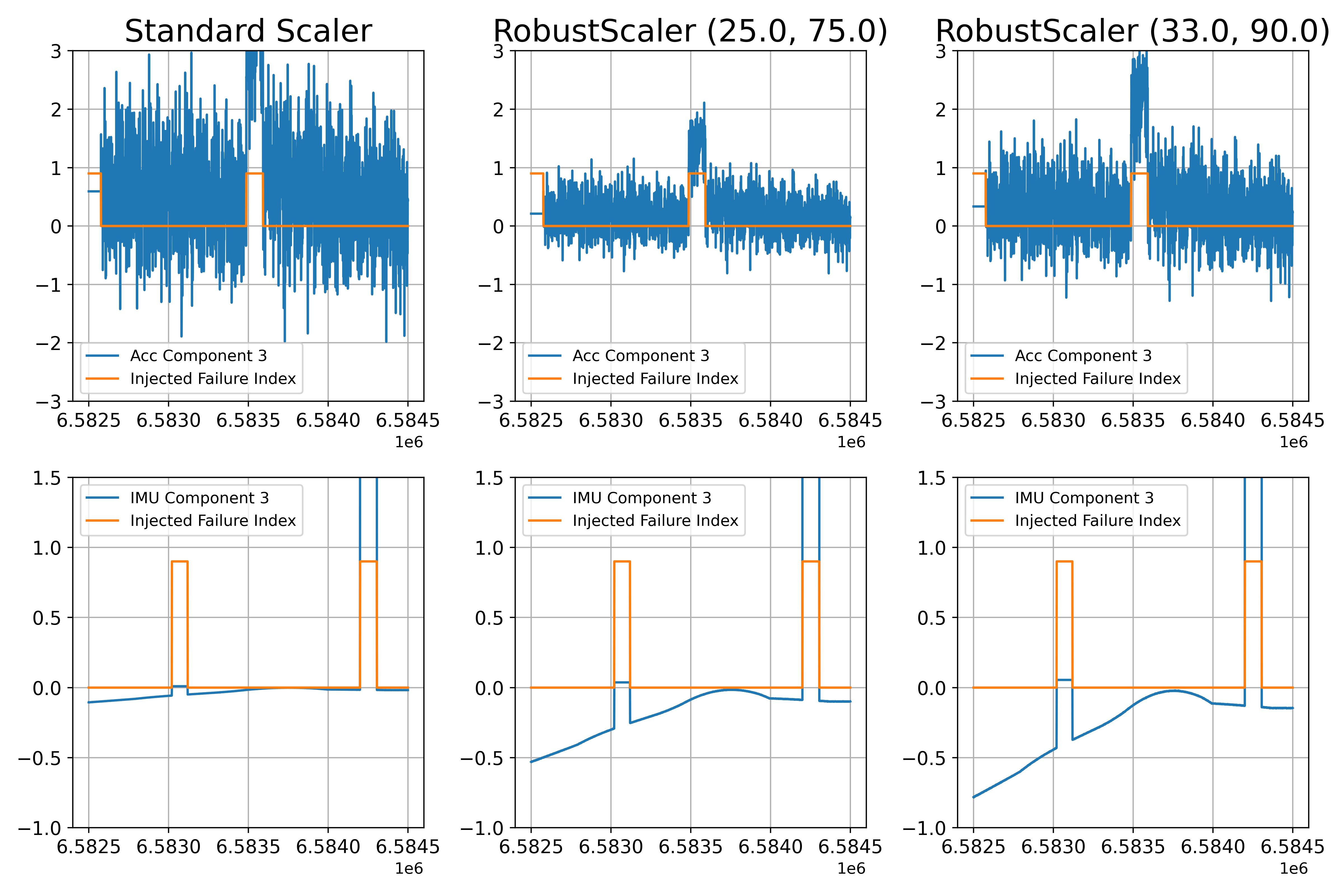}
    \caption{Comparison of Different Scalers Setups}
    \label{fig:scalers}
\end{figure*}

Special attention has been given in this work to proper data scaling as it plays a particularly relevant role in this study, besides being a well-recognized practice in AI literature (\cite{raschka2019python}).\\
Due to the presence of stuck at random value faults, the dataset can extend far beyond the nominal operational range of the accelerometer and IMU signals, ultimately approaching infinity. Hence, the resulting dataset is characterized by a large portion of data in proximity of the nominal range of the signal, with a reduced percentage of outliers assuming different random values. This spread of the dataset over a wide range of values is generally leading to deterioration of the Neural Network detection performances, unless a proper scaling is applied. \\
A comparison of different scalers from the \textit{scikit-learn} Python library was carried out with the objective of filtering the outliers from the nominal range of the signal, trying to provide a clear separation between the two and consequently enhance the fault detection. At the same time, an excessive shrink of the unfaulty signal shall be avoided, because it may deteriorate the performances of the Network in recognizing faults within its nominal range, especially stuck at last values.\\
A simple standardization of the entire dataset distribution would lead to excessive flattening of the nominal-range data near the zero line, as all data are centered and scaled based on their standard deviation, which is spoilt by the presence of infinite-like values. Hence, although a clear separation between the infinite-like values and the nominal is maintained, the stuck value cases other than these extreme values and the nominal signal are bounded within a range of a few orders of magnitude, leading to a consistent drop of the capability to distinguish one from the other, especially when noise is involved. To address this issue, \textit{scikit-learn}'s \textit{RobustScaler} has been selected to especially exclude infinite-value faults from the scaling computation. This scaler disregards a portion of the dataset when computing scaling factors by only keeping a specific quantile range, which means a percentage of the sorted data. \\
In the \textit{RobustScaler}, setting an upper bound to the quantile range is useful to exclude infinite-like values from computation, but may also help with those faults where the stuck value slightly deviates from the nominal range. This consideration applies equally to both the signal and its derivative, which is also a feature in the input of the neural network, as detailed in Section \ref{subs:cnn}.\\
Similar reasoning applies to the lower bound of the quantile range. Excluding the lowest values from scaling is particularly beneficial in the derivative of the signal, which sharply drops close to zero when a stuck value occurs, and thus can be adequately distinguished after the scaling stage. \\
Figure \ref{fig:scalers} presents a comparison between three scaler setups. In the first column, the distribution of the dataset is fully standardized, especially clarifying in the IMU case the shrink of the nominal range with respect to the random values. The second column represent the default setting of \textit{RobustScaler} from \textit{scikit-learn}, which shows a consistent improvement with respect to the previous case, bounding the Accelerometer signal in the $(-1,1)$ range, widening the nominal range of the IMU signal with respect to the outliers, and enhancing the separation of the faults in the two sensors from the rest of the data. The last column utilizes the quantile range obtained after the optimization described in Section \ref{subs:opt}. The selection of the quantile range as a hyperparameter reflects the crucial role of the scaling in this particular use-case to maximize the detection performances, leading to the decision to leave to the optimizer the task to provide the optimal value of this range, more than choosing it by trial and error.  

\subsection{Convolutional Neural Network} \label{subs:cnn}
\begin{figure}[h!]
    \centering
    \includegraphics[width=\textwidth]{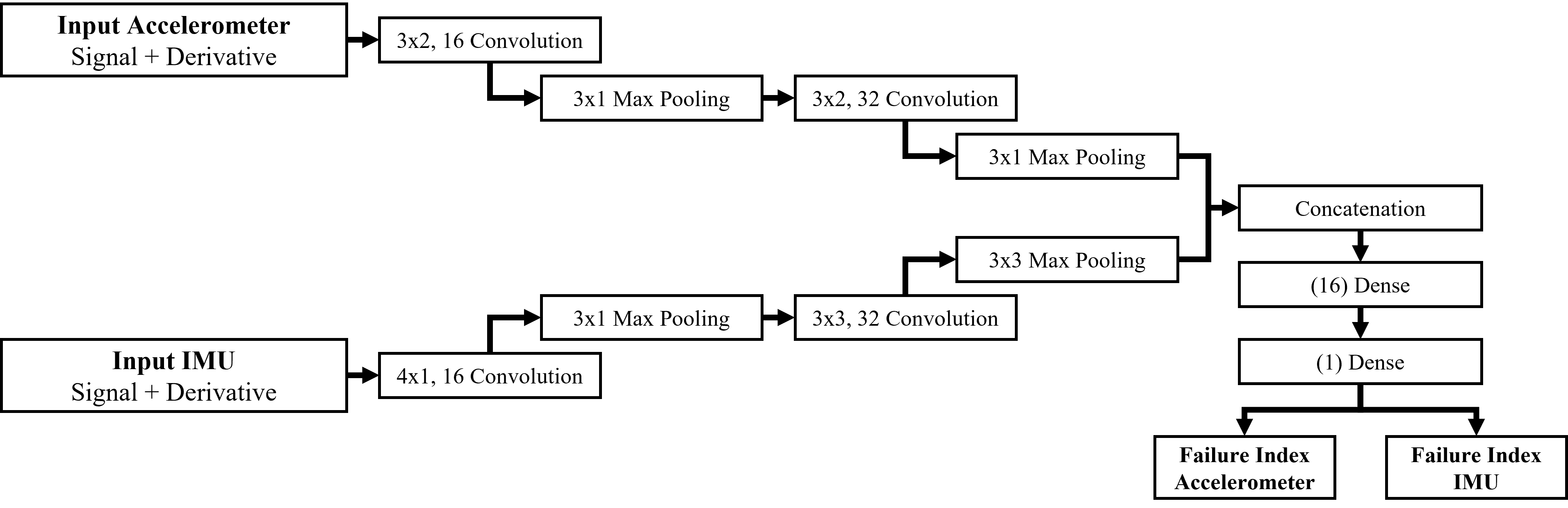}
    \caption{Structure of the Multi-Channel CNN}
    \label{fig:cnn_struct}
\end{figure}

% \subsubsection{Motivation}
In the specific FDI task addressed in this work, a CNN is employed for binary classification of input data into "fault" and "no fault". This decision was favored over its multi-class counterpart because, from the perspective of FDIR, there is no distinction among the recovery actions associated with the nature of the injected stuck value cases. However, the classification task must be multi-target, meaning that fault/no fault predictions must be obtained for both sensors under analysis. This is crucial for understanding where the fault is occurring and directing the recovery to the correct sensor.\\
The network structure must consider the interconnection between the signals of the two sensors, making it necessary to include a joint signal processing step. This ensures that complex scenarios requiring a comparison of the two signals to distinguish a fault are appropriately handled. For instance, a straight flight may result in a flat IMU signal, which could be misinterpreted as a stuck value if considered alone without cross-referencing the accelerometer output, which would instead be changing due to local gravity variations.\\
Inspired by the architecture proposed in \cite{cui2016multi}, this work proposes a multi-channel convolutional network where the input time series come from the two mentioned sensors, departing from the original idea of \cite{cui2016multi}, who proposed different manipulations of the very same time series as input. The output is instead made by two different failure indices, again corresponding to the two sensors in analysis (Figure \ref{fig:cnn_struct}).

The neural network takes input from a sliding window applied to the sensor signal, continually updating with new samples received from the simulator interface. This window comprises the current sample along with a fixed-length history of past samples, aiming to detect the evolution of a stuck value, which typically requires multiple samples to realize over time. In addition, the input window consists of six features: three representing the signal itself and three representing the derivative of the signal, computed at each time sample. The derivative addition aims to improve the recognition of stuck values within the nominal signal as it presents distinctive features in the presence of these faults. Specifically, the derivative tends to approach approximately zero whenever the fault starts: a precise zero value is obtained in the case of noise absence, while a more oscillating value is observed when noise is added. Also, specific spikes in the derivative can be observed for stuck at random values, whenever the signal jumps to a very different value in the frame of two subsequent samples.\\

The network structure is meant to emphasize interpretability, allowing for insights into which layer or hyperparameter influences specific output behaviors. On the one hand, each convolutional stage employs 1-dimensional convolutions, analyzing each feature independently, reflecting the independence of each component of the signal and the manifestation of the fault on different axes. On the other hand, the distinction between separate channels and a joint one enables the network to learn features from individual signals independently and capture connections between in a later stage. This approach facilitates the separate feature extraction from each signal in the initial stages and joint classification in subsequent layers following concatenation. Consequently, if the network underperforms on a particular sensor, tuning parameters in the associated branch suffices. Conversely, if the network struggles to detect faults where analysis of a single signal is insufficient, adjustments are made to the joint convolutional layer.

\subsection{Integration with onboard FDIR } \label{subs:integration}
The requirement of integrating the AI side-by-side with the PUS chain, coming from the project, led this study to consider different integration possibilities, which can eventually be linked to two main categories, either processing the AI failure index as an Event or as a pMon. The main difference between the two is that in the first case a single failure prediction from the AI is straightly translated to a recovery action, while in the second case the same prediction is first monitored as a pMon having its associated fMon, which eventually triggers the Event and the recovery afterwards.\\
In this specific case, the AI output has been decided to be processed as a single pMon with a single associated functional monitoring, with the expected value of the pMon set to zero. Since the output index is a Boolean value, the pMon is triggered whenever the AI output predicts a non-zero value (failure), and the associated fMon is triggered accordingly. \\
This approach offers several advantages. First, it enhances the robustness of the method to outliers compared to directly linking the prediction to event triggering. This is because it incorporates the persistency check between the failure prediction and event triggering. Additionally, it allows greater flexibility in implementation, particularly regarding the infrastructure surrounding the management of pMons and fMons. Indeed, the disabling of a pMon or fMon is a mechanism already embedded in the PUS-based FDIR (\cite{ECSS-E-ST-70-41C}), which can be reused for the AI output pMon, therefore enhancing its dependability as it is relying on well-established onboard procedures.

\section{Performances Evaluation} \label{sec:perf_ev}

Building on the research of \cite{nalepa2022evaluating} and \cite{mathonsi2022multivariate}, this study shares the objective of evaluating the CNN's ability to detect intervals rather than individual samples. However, unlike those works, this research does not explore intervals within the context of a signal reconstruction architecture. Instead, it addresses the classification of entire faulty intervals. Evaluating whether the CNN correctly classifies an interval is crucial, especially concerning the triggering of reactions in onboard FDIR through the mechanism of \textit{persistency} of positive failure predictions (see Section \ref{sec:intro}). In this context, the developed set of metrics serves to assess both the capability of the AI algorithm to detect faults (\textit{Detection Metrics}) and the effects of this detection on the onboard FDIR subsystem (\textit{System Metrics}).

\subsection{Detection Performances}
The Detection Metrics only assess the performance of the algorithm. Effectively substituting the classic \textit{precision, recall} and \textit{F1-score} metrics used to evaluate classification tasks, the set of metrics proposed in this work is meant to evaluate the capability of an algorithm to detect faults by measuring case-specific quantities that are more representative in the specific AI-based onboard FDIR framework.\\
Before properly defining the set of metrics, a few initial definitions need to be given.\\
The labels vector $y$ and the predicted labels vector $y_{pred}$ are divided into intervals of True (fault) and False (no fault) values. These intervals are referred to as \textit{faults} or \textit{fault intervals}. A \textit{Missed Fault} occurs when there is a complete interval of disagreement between $y$ and $y_{pred}$, meaning that the algorithm misses the whole fault. On the other hand, if the algorithm detects a fault with a positive prediction immediately after an interval of disagreement between $y$ and $y_{pred}$, it is referred to as a \textit{Delay} in prediction. Situations where $y_{pred}$ shows oscillations in the prediction while $y$ is consistently faulty are called \textit{Uncertain Prediction}. According to the given definitions, the Detection Metrics are presented in Table \ref{tab:det_metr}.
\begin{table*}[hbt!]
\centering
\caption{\label{tab:det_metr} Detection Metrics}
\begin{tabular}{|c|c|c| >{\centering\let\newline\\\arraybackslash\hspace{0pt}}m{4.5 cm}|}
\hline
    Metric                      & Type          & Objective       & Description \\ \hline 
    Missed Faults Score & Index         & $\rightarrow 0$ & Percentage of Missed Faults over total number of faults in the dataset\\\hline
    Prediction Delays           & Distribution  & Shrink          & Distribution of the Delays duration over the predicted fault intervals\\\hline
    Uncertain Prediction Score      & Index         & $\rightarrow 0$ & Percentage of faults with uncertain prediction over total number of faults in the dataset\\\hline
    False Positives Duration    & Distribution  & Shrink          & Distribution of the duration of the predicted false positives\\\hline
\end{tabular}
\end{table*}

\subsection{System Performances} \label{subs:sys_perf}
Related to the integration of the AI algorithm in a much broader system (i.e. the rest of the spacecraft), the system performances are related to the capability of impacting the components they interface with. More specifically, the AI-based FDIR is meant to exchange telemetry and telecommands with e.g. the OBC, the sensors themselves, etc. From some of these components it only receives data, from the other it establishes a bilateral link and is capable of triggering recovery actions to react to the detected failures. Therefore, the definition of System Metrics is tightly bounded to the integration strategy of the AI algorithm inside the PUS-based onboard FDIR subsystem.\\
In order to evaluate the performances of the architecture as defined in Section \ref{subs:integration}, a redefinition of the well-established concepts of Precision and Recall was carried out, starting from True Positives (TP), False Positives (FP) and False Negatives (FN), which are now reinterpreted in terms of the recovery actions triggered by the AI output within onboard FDIR. On one side, a predicted fault longer than the persistency time frame and with no real fault subtended is a FP because it triggers an unnecessary recovery action. Conversely, a real fault where no failure prediction is present constitutes a FN because it triggers no reaction. Besides these two extreme cases, it can also happen that a failure prediction with a real fault subtended does not meet the persistency duration requirement to trigger a reaction, therefore it has no impact on the system level and does not take part to the System Metrics computation. This case is particularly present in situations where the model is not well optimized, leading to high prediction delays which push the fault detection too late after the actual start, or leading to fragmented failure prediction within the same real faults where all the single fragmented prediction intervals are not able to reach the persistency threshold. Table \ref{tab:sys_metr} summarizes all the defined System Metrics.
\begin{table*}[hbt!]
\centering
\caption{\label{tab:sys_metr} System Metrics}
\begin{tabular}{|c|c|c| >{\centering\let\newline\\\arraybackslash\hspace{0pt}}m{4.5 cm}|}
\hline
    Metric                      & Type          & Objective       & Description \\ \hline
    Reaction Precision Score         & Index         & $\rightarrow 1$ & $\frac{TP}{TP+FP}$\\\hline
    Reaction Recall Score            & Index         & $\rightarrow 1$ & $\frac{TP}{TP+FN}$\\\hline
    False Positives Percentage       & Index         & $\rightarrow 1$ & Percentage of False Positive Reactions over theoretical number of triggered reactions in the dataset \\ \hline
\end{tabular}
\end{table*}

\begin{figure}[h!]
    \centering
    \includegraphics[width=0.6\textwidth]{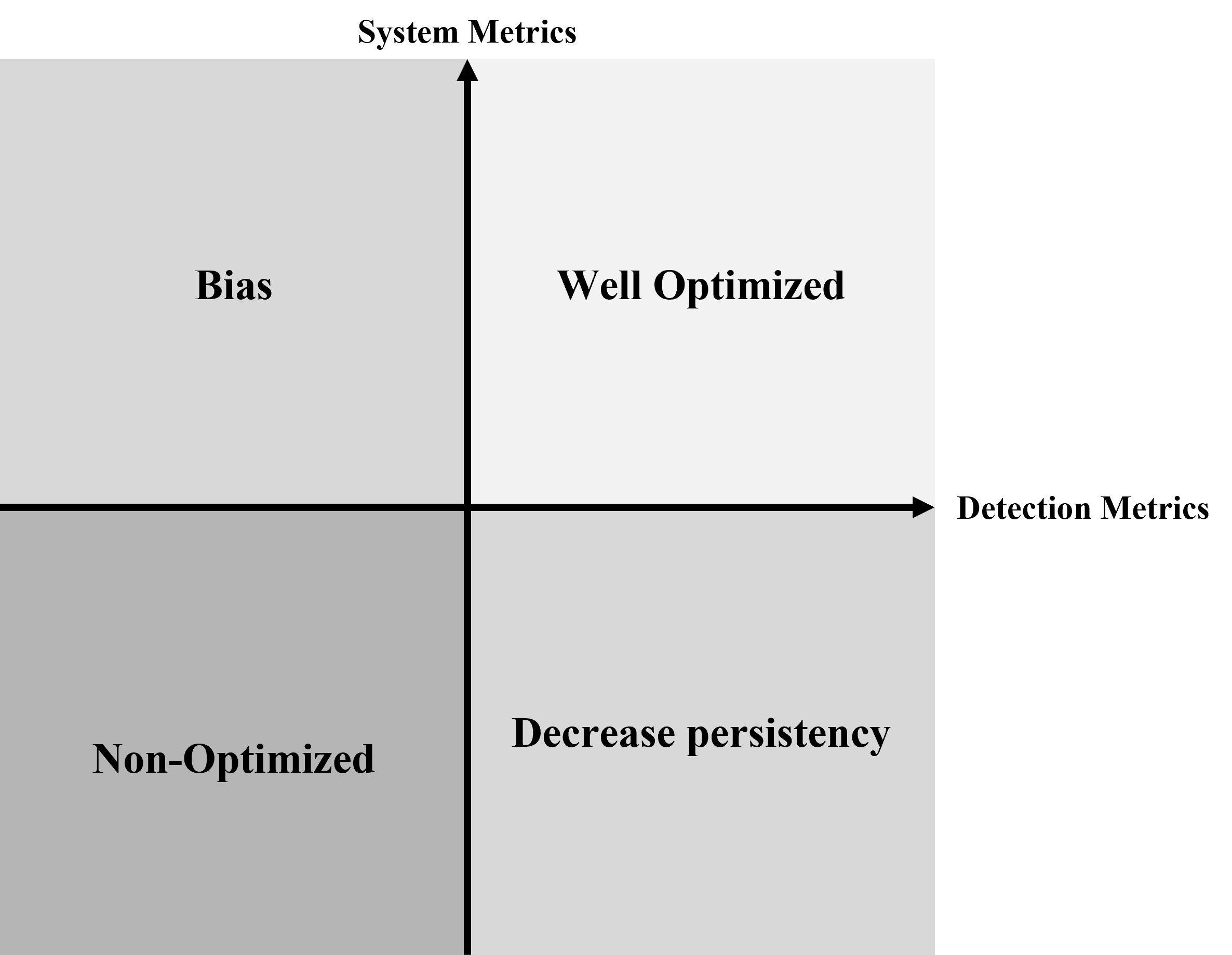}
    \caption{Correlation Scheme of Detection and System Metrics}
    \label{fig:dmsm_corr}
\end{figure}

It is important to note how each of the metrics defined in Table \ref{tab:sys_metr} depends on the choice of the persistency value, which is the crucial parameter to determine the triggering of a recovery action. The tuning of the persistency value is therefore the strategy to get the maximum from the System Metrics, which show a certain correlation with the Detection Metrics, as long as they reach an acceptable value. More specifically, an algorithm which is capable of reaching high detection performances likely has also good system performances, because good detection means that most of the faults are detected and the prediction is also smooth, allowing a more accurate reaction triggering based on the specified persistency value. In this best-case scenario, the two set of metrics are strongly correlated. However, taking into account the presence of outliers and false positives in general, delays and so forth, increased robustness to adverse reaction triggering is given by the persistency tuning. An appropriate tuning can first get rid of single outlier intervals lasting a few simulation samples, but also supply to those cases of uncertain prediction where the fragmentation of the prediction can be faced only by an appropriate persistency value. These considerations allow to pivot around non-perfect system performances and obtain the target results, being also an example of how the System Metrics are partially independent from the Detection Metrics. A summary of the correlation cases of Detection and System Metrics is presented in Figure \ref{fig:dmsm_corr}.

\subsection{Optimization} \label{subs:opt}
To understand how Detection and System Metrics are utilized for evaluating a specific AI-based FDIR solution, it is crucial to emphasize the underlying requirements. From the perspective of the FDIR subsystem, minimizing false positive-triggered reactions is paramount, as these could affect mission availability, potentially leading to the activation of safe mode. Conversely, while detecting faults is important, it is less critical compared to minimizing false positives, given the nature of the examined scenario. Indeed stuck values can always be indirectly identified by conventional FDIR methods through the propagation of their consequences over time. Hence, even if the AI is not perfect in detecting these faults, they can eventually be identified by PUS as time progresses. This mechanism is remarkable from a system engineering point of view because, since it is clear that an AI-based FDIR solution is not enough dependable (rather experimental) to be applied to an operational scenario, the stuck value faults provide a framework to prove the effectiveness of the AI on a critical functionality, without loading it with critical decision power.\\

The optimization of the specific AI-based FDIR solution has to focus on the System Metrics, which give a quantitative shape to the requirements mentioned above. However, the Detection Metrics also have to enter the optimization process by being used to double check if the value reached in the system performances is realistic or is subject to bias. Algorithm \ref{algo:full_opt} shows the complete optimization procedure.
\begin{algorithm}
	\caption{System Metrics Optimization}
        \label{algo:full_opt}
	\begin{algorithmic}[1]
		\State Start from an Established CNN Architecture
		\For {Different Hyperparameters Sets}
            \For {Different Quantile Range of Scaling}
            \For {Different Persistency Values}
%            \State Compute the Detection Metrics
		\State Compute the System Metrics 
		\State Compute the Objective Function (Equation \ref{eq:f_beta})
		\EndFor
            \EndFor
            \EndFor
            \State {Retrieve Optimal Parameters Value}
            \State {Tune the Persistency to Optimize the System Metrics}
		  \For {Persistency Values in an Interval Around the Optimized Value}
		\State Compute the System Metrics 
		\State Check Requirements Compliance
		\EndFor
            \State Double-check Detection Metrics
            \State {Freeze Persistency Value and Other Hyperparameters}
	\end{algorithmic} 
\end{algorithm} 

The introduction of persistency in the hyperparameter tuning loop is related to the need to compute and optimize the System Metrics, but is not conceptually against the independence of the System Metrics from the Detection Metrics. Indeed, the independence embedded in the design will be reflected in the optimization loop with the persistency value being optimized on its own, not depending on the rest of the hyperparameters, therefore making identical the process of optimizing the persistency aside the main loop or inside it. In addition, once the algorithm has given the optimized persistency value as output, the task of tuning the parameter for maximizing the system performances is already embedded and may eventually need minor adjustment based on the specific requirements.\\
Concerning the last step of Algorithm \ref{algo:full_opt}, the need to double-check the Detection Metrics at the end of the whole process arises from the idea that biases in metrics could occur if the algorithm excels in detecting long-lasting faults but struggles with shorter ones or sparse outliers. The optimization process may then lead to a high value of persistency following the long-lasting well-detected faults and even reach very good performances, while a check of the Detection Metrics would potentially show a high rate of Delays, False Positives, and Missed Faults. The check of the Detection Metrics is therefore needed in this case to identify this bias, before making considerations on its criticality. This specific consideration can be seen as an argument against putting the persistency inside the tuning loop in order to avoid the mentioned bias. However, this choice would have prevent the use of the System Metrics as optimization target, which need the persistency to be computed, leading to a conceptual inconsistency.\\

Based on the importance that the \textit{precision} acquires with respect to \textit{recall} from the system perspective, the objective function is chosen as a generalized F-beta score, shown in Equation \ref{eq:f_beta}.
\begin{equation} \label{eq:f_beta}
    F_{\beta} =  \sum_{i=1}^{n_{out}} w_{i} \, \frac{ (1+\beta^{2}) \cdot Pr_{i} \cdot Re_{i} }{ (\beta^{2} \cdot Pr_{i}) + Re_{i} } 
\end{equation}

In Equation (\ref{eq:f_beta}), the index $i$ ranges from 1 to the number of outputs in the Neural Network, in this case corresponding to the two sensors considered. $Pr$ and $Re$ indicate \textit{precision} and \textit{recall} respectively. $w_{i}$ and $\beta$ are two weighting factors, with the former weighting the contribution of the single Network output to the overall score, while the latter balancing the importance of precision and recall in the single terms composing the score. \\
For the specific objective function employed in this work, $w_{i}$ and $\beta$ have been set to 0.5, taking into account the equal weight of the two sensors on the overall result and the already mentioned importance of precision over recall.

\subsection{Results} \label{subs:res}

The training of the Neural Network occurs in a supervised way employing labels available from the fault injection stage. Due to the binary classification task, the \textit{binary crossentropy} loss function is employed, together with \textit{Adam} optimizer with learning rate subject to optimization, and an \textit{EarlyStopping} callback is set with patience 5. The batch size is 32768.
The rest of the hyperparameters are subject to the optimization process detailed in Algorithm \ref{algo:full_opt} and are shown in Table \ref{tab:hp}.
\begin{table*}[hbt!]
\centering
\caption{\label{tab:hp} CNN Hyperparameters}
\begin{tabular}{|l|l|}
\hline
    Hyperparameter                  & Value                     \\\hline
    Learning Rate                   & 0.0006507411205516692     \\\hline    
    Training Epochs                 & 32                        \\\hline
    Quantile Range Lower Bound (\%) & 33                        \\\hline
    Quantile Range Upper Bound (\%) & 90                        \\\hline
    Windows Length (samples)        & 180                       \\\hline      
    Persistency (samples)           & 27                        \\\hline
\end{tabular}
\end{table*}

It is remarkable to notice that the particular \textit{Windows Length} obtained after the optimization is smaller than the \textit{Distance Between Subsequent Faults} (Table \ref{tab:fail_inj_par}). The choice of the latter has been specifically done to ensure that each window would include maximum one fault, while including sufficient samples to allow the CNN to gather significant information on the stuck value eventually present. In other words, the network needs to see sufficient samples history to recognize a fault, if present, but cannot see windows where multiple faults are present, since this situation is not realistic. This consideration is equally valid for both sensors. Hence, the specific value chosen for the \textit{Distance Between Subsequent Faults} in Table \ref{tab:fail_inj_par} allows the optimizer to iterate in a significant range of windows length values, seeking optimal performances while ensuring the presence of one fault at a time.\\
The resulting Detection and System Metrics, evaluated on a validation dataset obtained by the very same Monte Carlo analysis described in Section \ref{subs:datagen}, are presented in Tables \ref{tab:det_metr_res} and \ref{tab:sys_metr_res}.
\begin{table*}[hbt!]
\centering
\caption{\label{tab:det_metr_res} Resulting Detection Metrics}
\begin{tabular}{|l|l|l|l|}
\hline
    Metric                          & Accelerometer & IMU       & Target            \\ \hline 
    Missed Faults Score             & 0.0503        & 0.1381    & $\rightarrow 0$   \\\hline
    Uncertain Predictions Score     & 0.1082        & 0.0699    & $\rightarrow 0$   \\
\hline
\end{tabular}
\end{table*}

\begin{figure}[h!]
    \centering
    \begin{subfigure}[b]{0.45\textwidth}
        \centering
        \includegraphics[width=\textwidth]{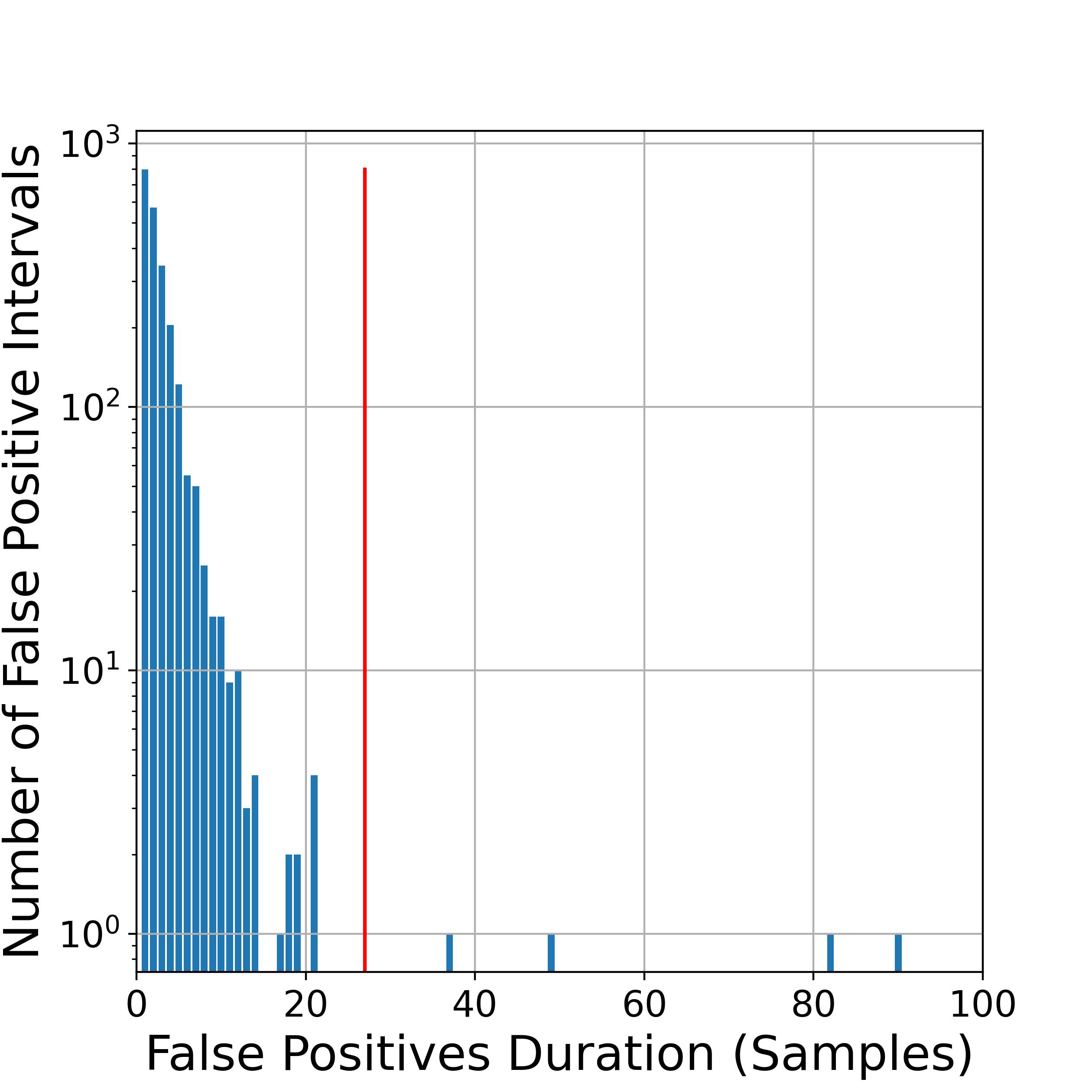}
    \end{subfigure}
    \hfill
    \begin{subfigure}[b]{0.45\textwidth}
        \centering
        \includegraphics[width=\textwidth]{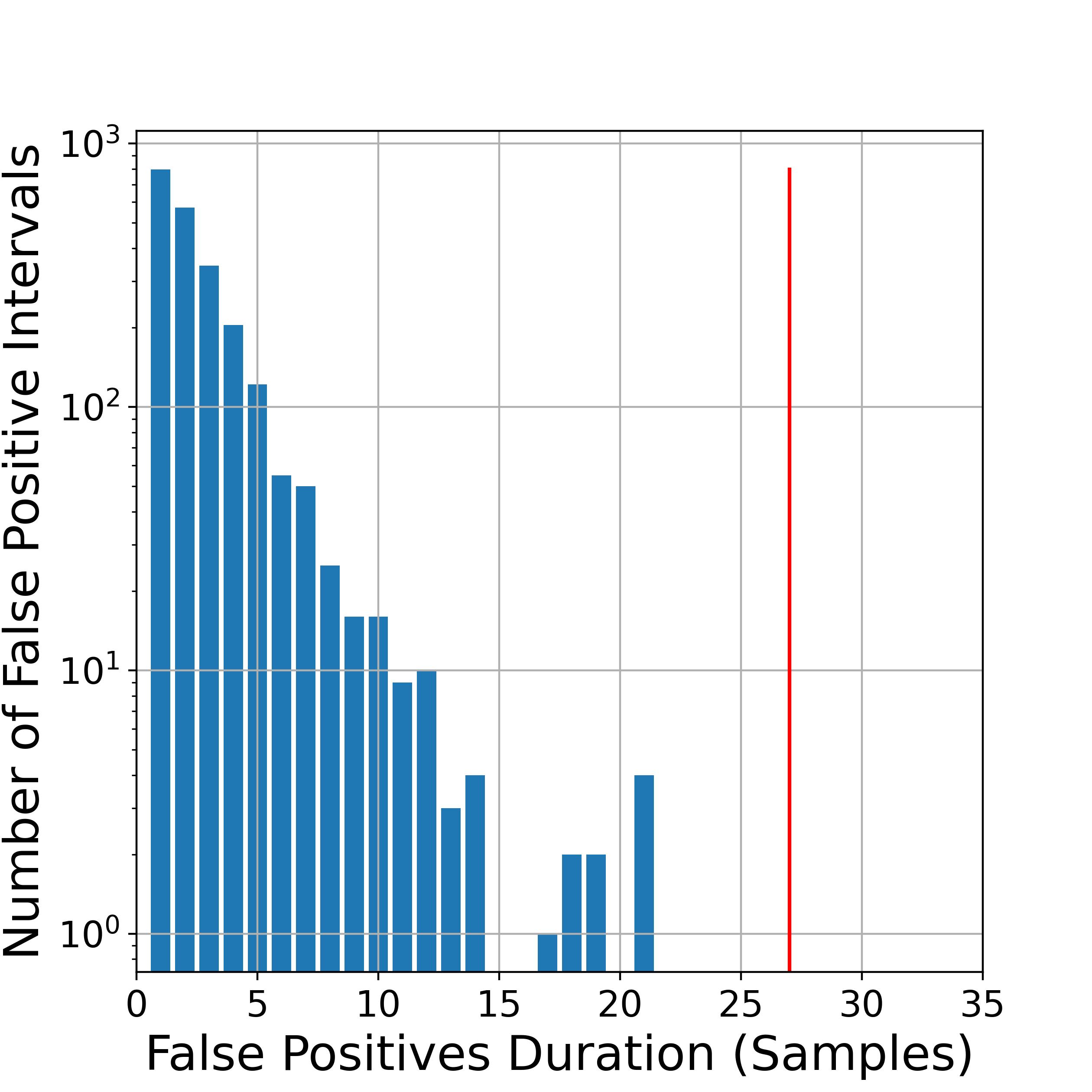}
    \end{subfigure}
    \caption{Distribution of the False Positives Intervals for the Accelerometer, persistency highlighted in red. Full duration range (left) and close-up (right)}
    \label{fig:false_pos_hist_acc}
\end{figure}

\begin{figure}[h!]
    \centering
    \begin{subfigure}[b]{0.45\textwidth}
        \centering
        \includegraphics[width=\textwidth]{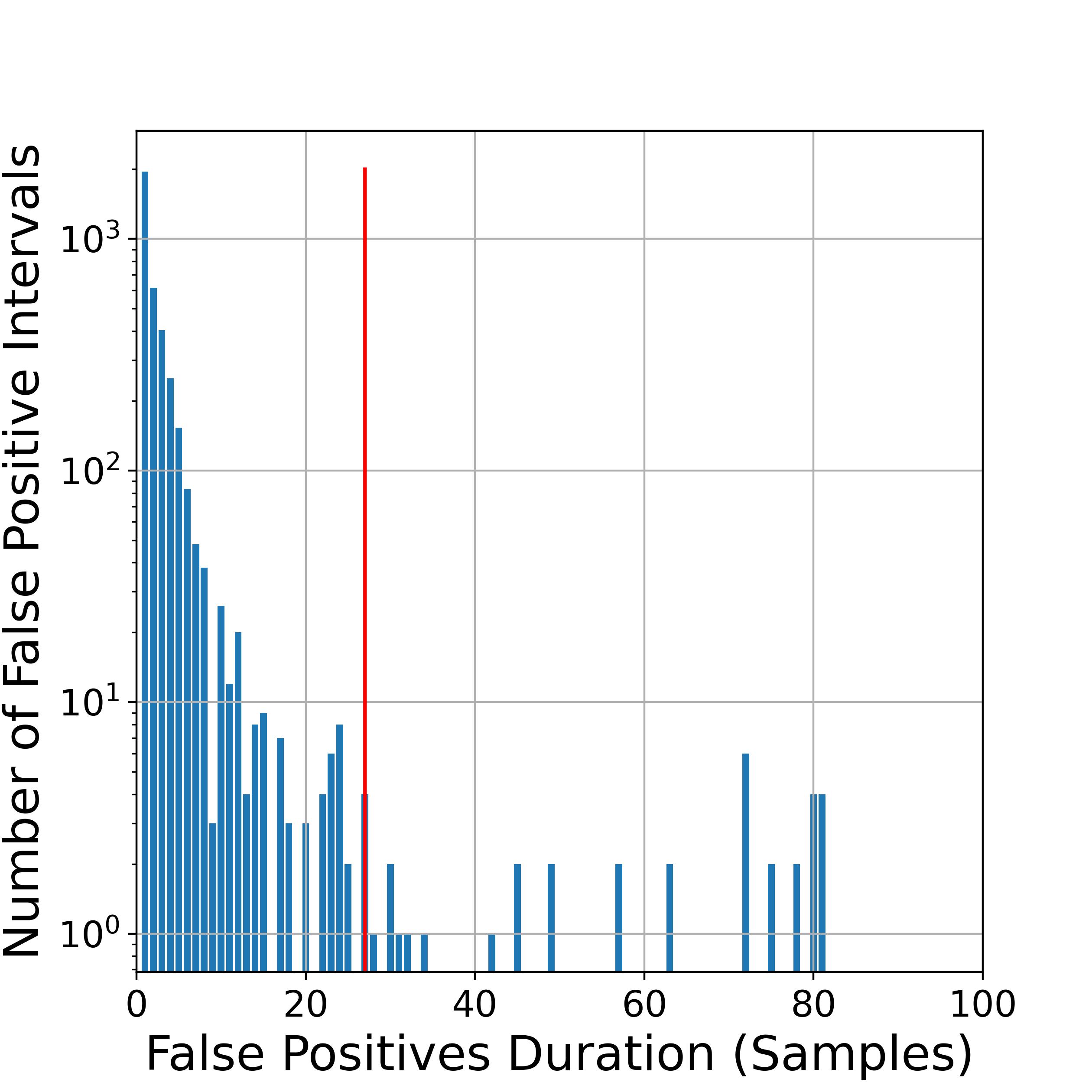}
    \end{subfigure}
    \hfill
    \begin{subfigure}[b]{0.45\textwidth}
        \centering
        \includegraphics[width=\textwidth]{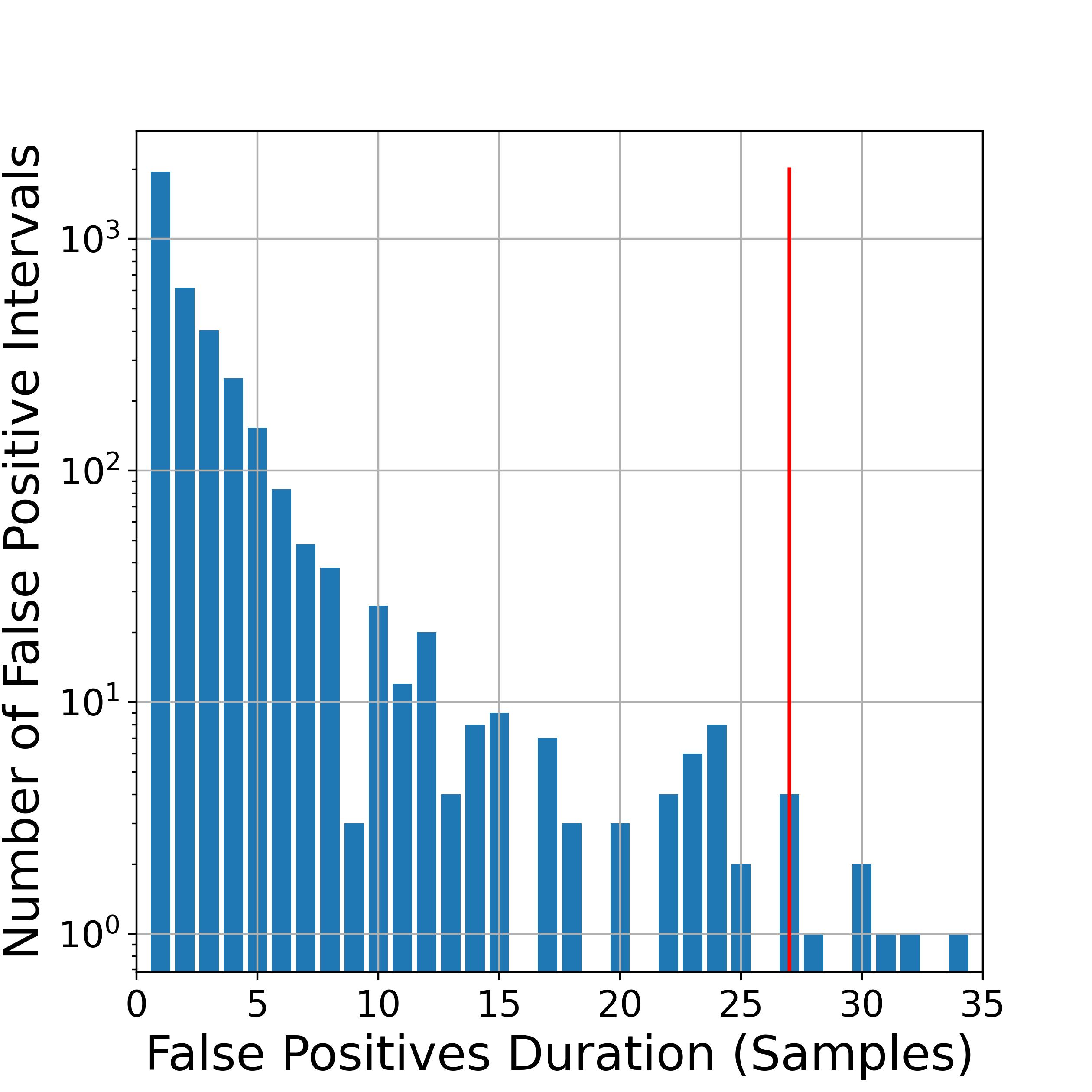}
    \end{subfigure}
    \caption{Distribution of the False Positives Intervals for the IMU, persistency highlighted in red. Full duration range (left) and close-up (right)}
    \label{fig:false_pos_hist_imu}
\end{figure}

\begin{table*}[hbt!]
\centering
\caption{\label{tab:sys_metr_res} Resulting System Metrics, $Persistency=27$ samples}
\begin{tabular}{|l|l|l|l|}
\hline
    Metric                          & Accelerometer & IMU       & Target            \\ \hline 
    Reaction Precision Score        & 0.99          & 0.99      & $\rightarrow 1$   \\\hline
    Reaction Recall Score           & 0.95          & 0.85      & $\rightarrow 1$   \\\hline
    False Positives Percentage      & 0.0006        & 0.0057    & $\rightarrow 0$   \\\hline
\end{tabular}
\end{table*}
\newpage
Figure \ref{fig:false_pos_hist_acc} and \ref{fig:false_pos_hist_imu} show the False Positives Duration distribution for the accelerometer and IMU case respectively, highlighting the value of 27 samples for the persistency.

Both the two distributions are skewed towards the left side of the persistency, indicating that this particular value effectively filters out most false positives, whether they are outliers or consistently long incorrectly classified intervals. Plus, such a value of the persistency poses itself below the lower bound of the injected fault duration, meaning that no injected fault is inherently neglected. In other words, the CNN is able to detect even the shorter injected faults. This outcome is not trivial, because if the algorithm had been not sufficiently expressive, the persistency could have been pushed back by the optimization algorithm to neglect shorter faults in favor of the longer ones, ultimately falling into the biases outlined in Section \ref{subs:sys_perf}.\\
Moreover, the two False Positives distributions present different spreading over the range from 30 samples to 110, which are the boundaries of the injected faults duration. The difference of the spreading in the two sensors is explained by the different performances achieved after the optimization procedure, with the accelerometer performing better than the IMU. However, the samples falling beyond the vertical line indicate that there is a number of False Positive intervals which is not filtered out by the persistency, causing the deterioration of the System Metrics, as shown in Table \ref{tab:det_metr_res}.\\

\begin{figure}[h!]
  \centering
  \begin{subfigure}{0.45\textwidth}
    \centering
    \includegraphics[width=\linewidth]{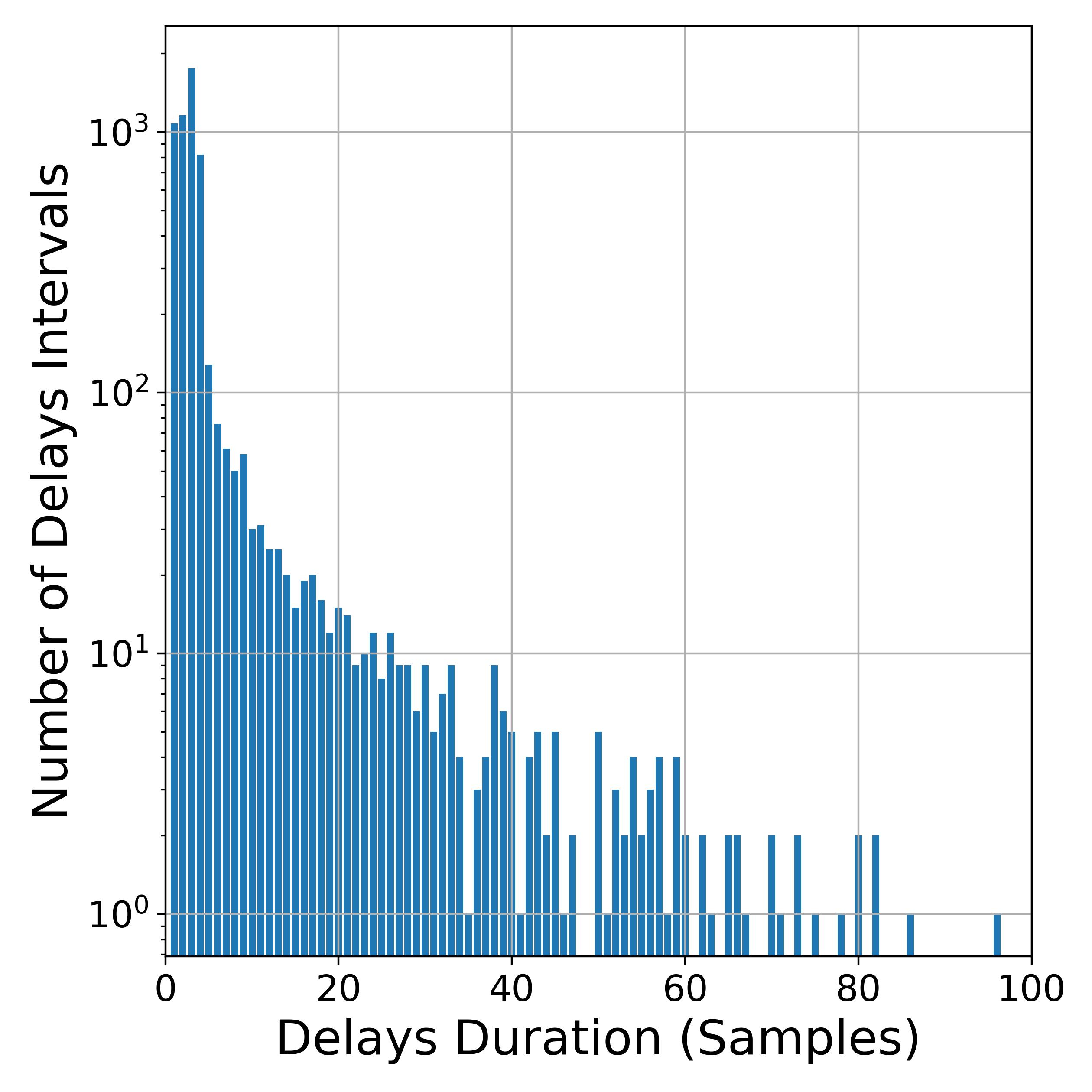}
  \end{subfigure}
  \hfill
  \begin{subfigure}{0.45\textwidth}
    \centering
    \includegraphics[width=\linewidth]{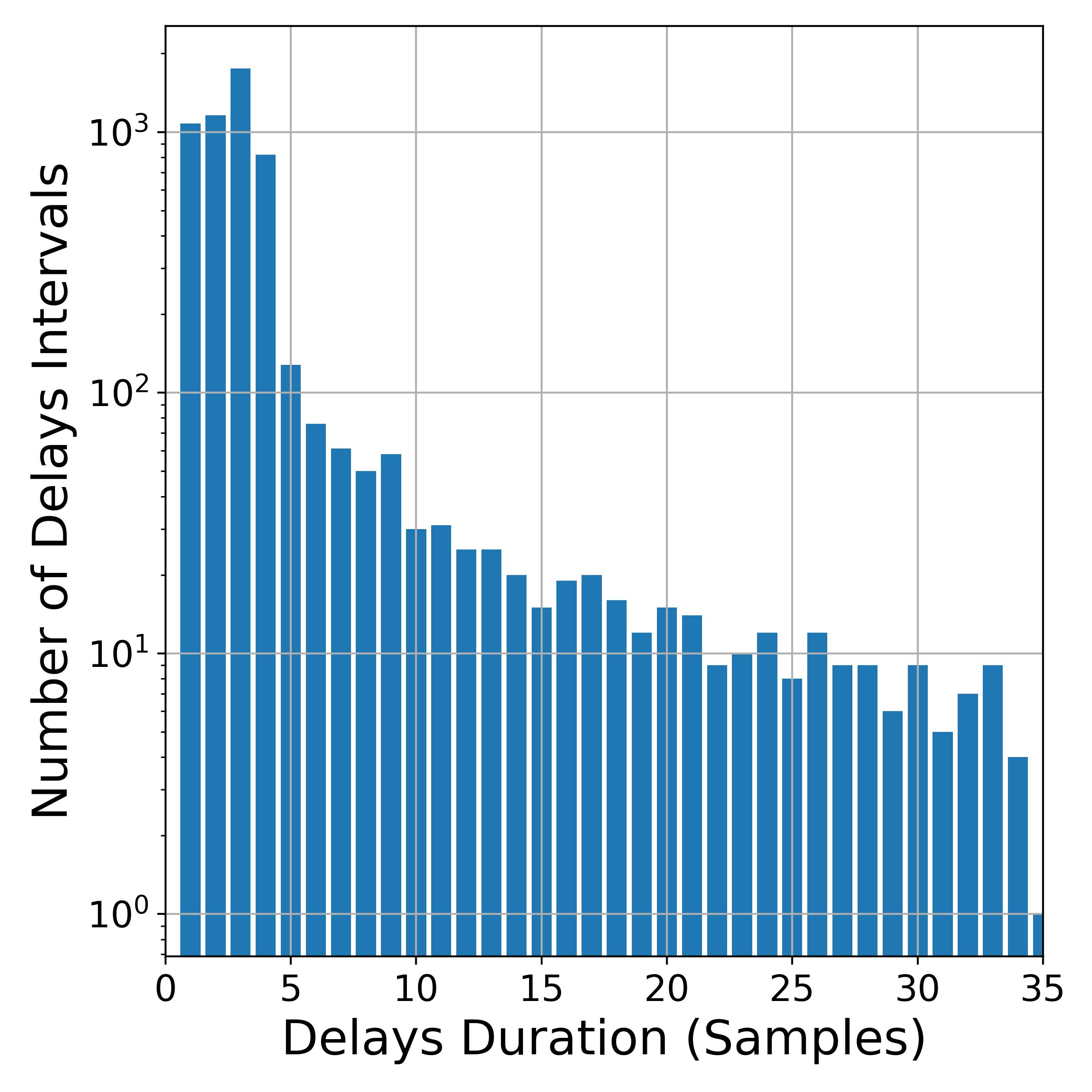}
  \end{subfigure}
  \caption{Prediction Delays Distribution for the Accelerometer case. Full duration range (left) and close-up (right)}
  \label{fig:delays_hist_acc}
\end{figure}

\begin{figure}[h!]
  \centering
  \begin{subfigure}{0.45\textwidth}
    \centering
    \includegraphics[width=\linewidth]{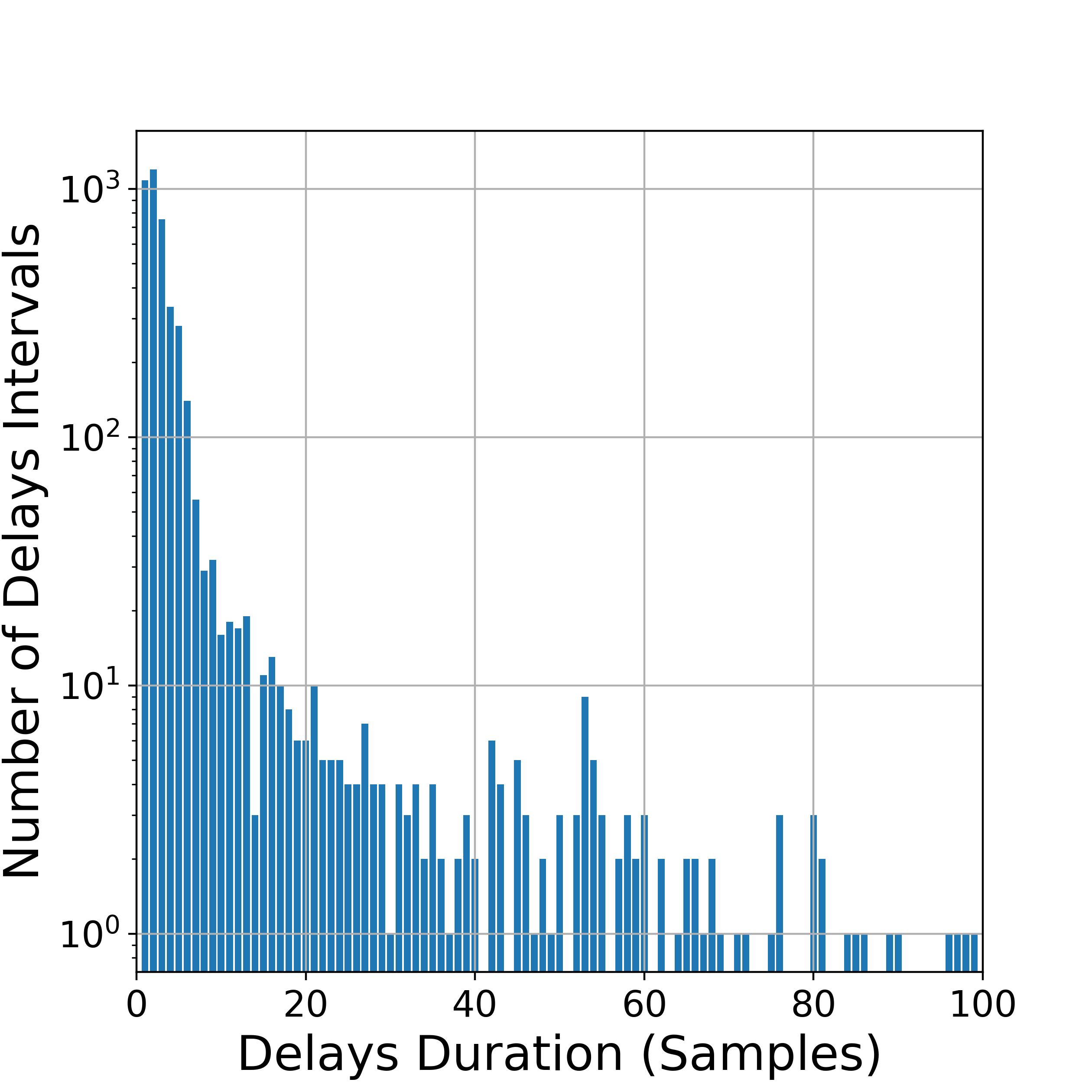}
  \end{subfigure}
  \hfill
  \begin{subfigure}{0.45\textwidth}
    \centering
    \includegraphics[width=\linewidth]{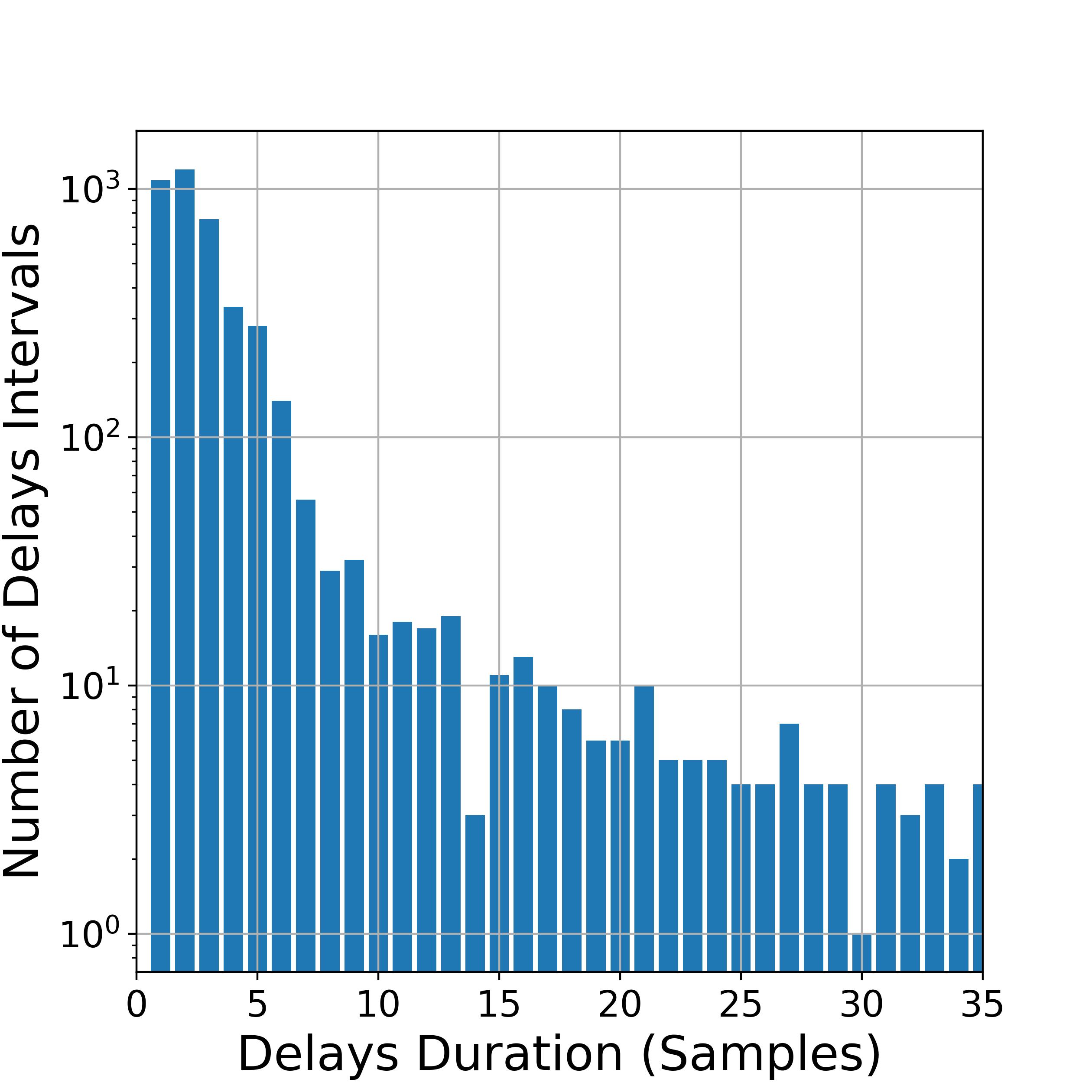}
  \end{subfigure}
  \caption{Prediction Delays Distribution for the IMU case. Full duration range (left) and close-up (right)}
  \label{fig:delays_hist_imu}
\end{figure}

Lastly, the Prediction Delays Distribution is shown in Figure \ref{fig:delays_hist_acc} and Figure \ref{fig:delays_hist_imu} for both the sensors. 

The two distributions namely present a concentration of values below 5 samples, with different shapes, corresponding to those correctly detected faults where the AI inherently needs a few sample to detect the stuck value as it progresses in time. Increasing the Delays Duration, the spreads of the plot over the entire duration range of the injected faults suggests the presence of cases where faults are detected after much of their progress, even too late to achieve a correct triggering of the recovery action. The objective of the optimization is in general to avoid these cases, which means shrinking the distribution of delays towards a small range of Duration, to enhance the detection of most of the faults and consequently improve the System Metrics. However, the presence of delays is inherent in the model and cannot be totally deleted. This fact, considering also the persistency mechanism, leads to the conclusion that the shortest detectable fault for the presented solution shall last a number of samples equal to \textit{minimum observed delay}+\textit{persistency}.

\section{Conclusions and Future Work} \label{sec:results}

The different stuck value faults considered in the simulation comprehensively represent the various realizations of this class of faults in realistic scenarios on a functional level, making the presented solution applicable on different FDIR levels. The employed simulation method also claims to be representative of realistic occurrences of stuck values, based on RAMS analysis, to enhance the learning of the CNN which can more easily adapt to a potential deployment. Since the dataset construction methodology, detailed in Section \ref{subs:datagen}, is focused on an early detection of the faults, it does not need to include realistic fault duration, thus enhancing the generalization of the detection to stuck values occurring at very different time scales, i.e. in very different components. The only limitation is in the end constituted by the earliest fault detection capability, expressed in the \textit{Prediction Delays} metric (Figure \ref{fig:delays_hist_acc} and Figure \ref{fig:delays_hist_imu}), which set the lower bound to the duration of the detectable faults. Clearly, this reasoning has to take into account the persistency value, which has to stay coherent with the targeted faults.\\
Based on the accelerometer and IMU of \textit{Astrone KI}, the proposed dataset has been subject to fault detection via a Convolutional Neural Network (Section \ref{subs:cnn}), which has proven its effectiveness in the diagnosis of the faults, achieving outstanding performances. The convolutions-based approach guarantees a partially interpretable algorithm which, in this particular case, allows a clear separation between the analysis of the single sensors and the joint-channel operations. Besides, it is particularly useful for the analysis of different time scales trends in time series, thus to address the stuck values faults presented in this work. Distinguishing a stuck derivative value behind the white noise to classify it is the most obvious situation, that the present analysis faced, where convolutions served the scope of recognizing trends at different time scales.\\
A set of case-specific metrics is then proposed to evaluate the CNN in its fault detection task. The design of the metrics focused on the real-time scenario, as well as on the interval-based nature of the faults. The latter is an element of innovation because it detaches from the classic point-wise conception of the fault classification, adapting to a problem which is very common in the field of anomaly detection in time series. Consequently, the developed metrics add a further layer of complexity to the network output processing, but makes it way more interpretable than classic evaluation approaches (i.e. \textit{F1-score}, \textit{Precision}, \textit{Recall}). \\

While the present research shows promising baseline performances on the \textit{Astrone KI} use-case, it is just meant to prove the feasibility of the approach to accomplish the specific stuck value fault detection task. A deployment of the presented CNN onboard a flying mission shall first pass through the definition of application-dependent requirements (e.g. Precision and False Positives Percentage). These performance criteria shall be meant to guide the tuning process, as described in Section \ref{subs:opt}, thus shall be defined \textit{a priori} based on the specific objectives of the mission, on the type of failure to be addressed and their required recovery actions. A first complement to the approach presented in this paper can be represented by a fine-tuning of $w_{i}$ and $\beta$ in Equation (\ref{eq:f_beta}), to obtain performances tailored to the specific needs. Plus, a sensitivity analysis over these parameters can also be an asset to evaluate the optimality of the obtained solution and enhance its robustness.\\
Concerning the qualification of the presented model for space applications, the deployment on space-representative hardware (e.g. FPGAs) is certainly important, aiming to validate the solution in realistic scenarios and identify any necessary adjustments or redesigns. This innovation would be eventually needed to prove the compatibility of the algorithm with field-specific edge devices in a real-time scenario, but also that it is able to achieve a similar performance as the theoretical one shown in the present work.\\
Similarly, another important step to the qualification for space is to test the robustness of the presented algorithm to different sensor models, coming from various producers in the space domain. This procedure would ensure the algorithm to fit various applications, enhancing its reusability across different missions and use-cases.\\

\section*{Acknowledgements}

The results presented in this paper have been achieved by the project Astrone - Increasing the Mobility of Small Body Probes, which has received funding from the German Federal Ministry for Economic Affairs and Energy (BMWi) under funding number “50 RA 2130A”. The consortium consists of Airbus Defence and Space GmbH, Astos Solutions GmbH, Institute of Automation (Technische Universität Dresden) and Institute of Flight Mechanics and Controls (Universität Stuttgart). Responsibility of the publication contents is with the publishing authors.

%% Bibliography
%% Author year style
% Bibliografia
\printbibliography % Stampa la bibliografia

\end{document}